\documentclass[aps,pra,twocolumn]{revtex4}
\usepackage{amsfonts}
\usepackage{bbding}
\usepackage{amssymb}
\usepackage{color}
\usepackage{amsmath}
\usepackage{graphicx}
\usepackage{subfigure}
\usepackage{txfonts}
\usepackage{bm}
\usepackage{ulem}
\usepackage[colorlinks,linkcolor=blue,citecolor=blue]{hyperref}

\begin{document}
\title{Nonuniform phases in the geometrically frustrated dissipative XYZ model}
\author{Xingli Li and Jiasen Jin}
\email{jsjin@dlut.edu.cn}
\affiliation{School of Physics, Dalian University of Technology, Dalian 116024, China}
\date{\today}

\begin{abstract}
We investigate the steady-state phase diagram of the dissipative spin-1/2 XYZ model on a two-dimensional triangular lattice, in which each site is coupled to a local environment. By means of cluster mean-field approximation, we find that the steady-state phases of the system are rather rich, in particular there exist various types of nonuniform antiferromagnetic phases due to the geometrical frustration. As the short-range correlations included in the analysis, the numerical results show that the oscillatory phase disappears while the triantiferromagnetic and biantiferromagnetic phases remain to exist in the thermodynamic limit. Moreover, the existence of the spin-density-wave phase, which is missed by the single-site mean-field analysis, is also revealed by the spin-structure factor.
\end{abstract}

\maketitle

\section{Introduction}
The collective behaviors in quantum many-body systems have attracted much attention in modern physics \cite{sachdev_book}. Usually, frustration leads to exotic properties of many-body system. The frustration refers to the presence of competing interactions that cannot be simultaneously satisfied. A typical example of frustrated system is the spin Ising model on a triangular lattice \cite{wannier1950}. Here the frustration stems from the geometry of lattice and thus named as geometric frustration. Although the nearest-neighboring interaction favors the spins being anti-aligned, the three spins on a triangle cannot be antiparallel. Due to the presence of frustration effect, the fluctuations of the ground state are enhanced giving rise to a large degeneracy. A large variety of intriguing quantum phases may emerge in frustrated system, such as spin glass \cite{sherrington1975}, spin liquid \cite{balents2010,zhou2017}, spin ice \cite{lee2002}, and supersolid \cite{jiang2009,zhang2011}. Experimentally, the geometric frustration has been realized in the system of trapped atomic ions \cite{kim2010} and optical lattice \cite{struck2011}.

However, a system always becomes open due to the unavoidable interactions to its surrounding environment \cite{breuer_book}. The system-environment coupling leads to the dissipative process of system and drives the system away from equilibrium. In the presence of dissipation, the dynamics of system is described by the master equation. Rather than the ground-state properties in the equilibrium many-body system, we are interested in the long-time limit properties of the open many-body system. In the long-time limit, the system may approach to an asymptotic steady state which corresponds to a fixed point in the phase space. The steady state can break the symmetry of the master equation and constitute novel phases that never appear in the equilibrium counterpart \cite{lee2011,lee2013,jin2013,leboite2014,weimer2015,schiro2016,landa2020a}. Alternatively, the system may end up in a limit cycle showing a periodic oscillation which intimately relates to the quantum time crystals \cite{gong2018,iemini2018,tucker2018,gambetta2019,landa2020b} and quantum synchronization \cite{ludwig2013,DT2018}. Recently, the steady-state phase transitions have been experimentally observed in circuit QED lattices \cite{fitzpatrick2017,collodo2019} and ensemble of Rydberg atoms \cite{carr2013}.

Essentially, the peculiar behavior of the steady state originates from the competition between the coherent dynamics governed by the many-body Hamiltonian and the dissipative process. In addition, it is found that the presence of frustration can indeed enrich the steady-state properties. For instance, the geometric frustration in the dissipative spin system leads to the various antiferromagnetism and chaotic dynamics \cite{qian2013};  The steady-state of frustrated driven-dissipative photonic cavities exhibit strongly correlated dark state \cite{rota2017a} and is analogous to the spin-liquid phase if the two-photon driving and losses are considered \cite{rota2019}.

In this work, we study the dissipative spin-1/2 XYZ Heisenberg model on a two-dimensional triangular lattice. By employing the state-of-the art numerical approaches, we explore the steady-state phases of this model. We notice that the same model in two-dimensional square lattice have been widely studied with diverse numerical approaches by focusing only on the steady-state ferromagnetic phase \cite{lee2013,jin2016,rota2017b,kshetrimayum2017,biella2018,rota2018,nagy2018,casteels2018,huybrechts2019}. By means of the numerical linked-cluster expansion, the dynamics and phase transition from the paramagnetic to ferromagnetic phases of the dissipative XYZ model on triangular lattice have been investigated \cite{qiao2020}. Here, we concentrate on the antiferromagnetic region where the geometric frustration plays crucial role in determining the steady-state phase. We uncover the various antiferromagnetic phases with different configurations of magnetization. Moreover, we also find the evidence of the spin-density-wave (SDW) phase.

The paper is organized as follows: In Sec.\ref{section:model}, we explain the frustrated dissipative spin-1/2 XYZ model on triangular lattice and the corresponding master equation that describes the evolution of the system. We also present the possible steady-state phases that may appear in the phase diagram. In Sec. \ref{subsection:Mean-fieldPhaseDiagram}, as a preliminary exploration, we implement the single-site mean-field (MF) analysis to the steady-state phase. The phases are characterized by the steady-state magnetization which are the stable solutions to the MF master equation. In Sec. \ref{section:beyond mean-field approximation}, we refine the MF results by the cluster mean-field (CMF) approaches in which the short-range correlations are systematically included as the size the of cluster increases. We check the existence of the ferromagnetic, antiferromagnetic and SDW phases in the thermodynamic limit through the Liouvillian gap and the spin-structure factor. We summarize in Sec.\ref{section:conclusions}.

\section{Model}
\label{section:model}

The model we consider here is a spin-1/2 XYZ system on a triangular lattice whose Hamiltonian is given by
 (setting $\hbar$ = 1 hereinafter),
\begin{equation}
H = \sum_{\langle j,k\rangle}(J_{x}\sigma^{x}_{j}\sigma^{x}_{k} + J_{y}\sigma^{y}_{j}\sigma^{y}_{k} + J_{z}\sigma^{z}_{j}\sigma^{z}_{k}),
\label{XYZHamiltonian}
\end{equation}
where $\sigma^{\alpha}_{j}$ $(\alpha = x,y,z)$ are the Pauli matrices on the $j$-th site and the sum of spin-spin interactions runs over the nearest neighboring sites $\langle j,k \rangle$. In addition, we assume that system is coupled to the Markovian environment. Effectively, each spin is subjected to a local dissipative channel that incoherently flips the spin down to $z$-direction.
%In the case of anisotropic coupling, the precession about an effective magnetic field induced its neighbors and the dissipation is responsible to the emergence of nontrivial magnetic order.
Therefore, the dynamics of the system is governed by the following master equation in the Lindblad form,
\begin{equation}
\frac{d\rho}{dt}={\cal L}[\rho]=-i[H,\rho] + \sum_{j}\mathcal{D}_{j}[\rho],
\label{eq:masterequation}
\end{equation}
where ${\cal L}$ is the Liouvillian superoperator and the local Lindbladian superoperator reads
\begin{equation}
\mathcal{D}_{j}[\rho]=\frac{\gamma}{2}\left[2\sigma^{-}_{j}\rho\sigma^{+}_{j}-\{\sigma^{+}_{j}\sigma^{-}_{j},\rho\}\right],
\label{eq:Lindbladian}
\end{equation}
with $\gamma$ denoting the decay rate and the operators $\sigma^{\pm}_{j}=(\sigma^{x}_{j}\pm i\sigma^{y}_{j})/2$ are the raising and lowering operators for the $j$-th spin, respectively. In deriving the Lindblad master equation, the Born-Markovian and the secular approximations are considered. These assumptions can be met in the experimental situations with circuit QED lattice \cite{fitzpatrick2017,collodo2019}. Recently, other type of master equation beyond Lindblad form have also been developed to study the steady-state properties of open quantum many-body systems \cite{lebreuilly2017,xu2019,reis2020}.

We are interested in the steady state of system which is given by the asymptotic solution to Eq. (\ref{eq:masterequation}) in the long-time limit $\rho_{\text{ss}} = \lim_{t \rightarrow\infty}{\rho(t)}$ (we use the subscript `ss' to denote steady-state). The steady-state property of the system is characterized by the expectation value of an appropriate operator $\langle O\rangle_{\text{ss}}=\text{tr}(O\rho_\text{ss})$.

The master equation in Eq. (\ref{eq:masterequation}) possesses a $\mathbb{Z}_{2}$ symmetry, which is invariant under a $\pi$-rotation along the $z$-axis $(\sigma^{x}_{i}\to-\sigma^{x}_{i},\sigma^{y}_{i}\rightarrow-\sigma^{y}_{i}, \forall i)$. In the thermodynamic limit, as the coupling constants changing, the steady state of the system may undergo a phase transition from the disordered paramagnetic (PM) phase, with vanishing magnetization in the $x$-$y$ plane, to ordered phase, with nonzero $\langle\sigma^x\rangle_{\text{ss}}$ and $\langle\sigma^y\rangle_{\text{ss}}$, corresponds to the spontaneously broken of the $\mathbb{Z}_{2}$ symmetry. In particular, the ordered phase can be uniform with identical nonzero $\langle\sigma^x\rangle_{\text{ss}}$ or $\langle\sigma^y\rangle_{\text{ss}}$ through the whole lattice, referred as ferromagnetic (FM) phase. Alternatively, it can be nonuniform with a spatial modulated magnetization of wavelength equal to or greater than the lattice constant, referred as antiferromagnetic phase or SDW phase, respectively. We stress that the anisotropic coupling $J_x\ne J_y$ is necessary to establish the ordered phases. Because the magnetization along the z-direction is conserved for $J_x=J_y$. Therefore, there is nothing to counteract the dissipation. The system will eventually approach to the PM phase with all the spin pointing down to the $z$-direction.

The emergence of the ordered phases can be understood as follows. Apart from the dissipation, each spin precesses about an effective magnetic field depending on the direction of its neighboring spin. If the precession is strong enough to counteract the dissipation, the spins will point away from the $z$-direction in the steady state indicating the emergence of ordered phases. Due to the geometric frustration of the triangular lattice, the neighboring spins on the triangles can not align antiparallel in the $x$-$y$ plane in the antiferromagnetic region. This leads to different types of antiferromagnetic orders in steady state as will be discussed in later sections.

We also note that, for $J_z=0$, the Hamiltonian in Eq. (\ref{XYZHamiltonian}) is reduced to the XY Hamiltonian. The steady-state phases of the dissipative XY model have been reported in Refs. \cite{joshi2013,wilson2016,maghrebi2016,marzolino2016,parmee2020}. In this paper, we concentrate on the case of $J_z/\gamma = 1$. The analysis on other values of $J_z \ne 0$ works similarly.

\section{Mean-field phase diagram}
\label{subsection:Mean-fieldPhaseDiagram}

\begin{figure}[!htbp]
\includegraphics[width=0.95\linewidth]{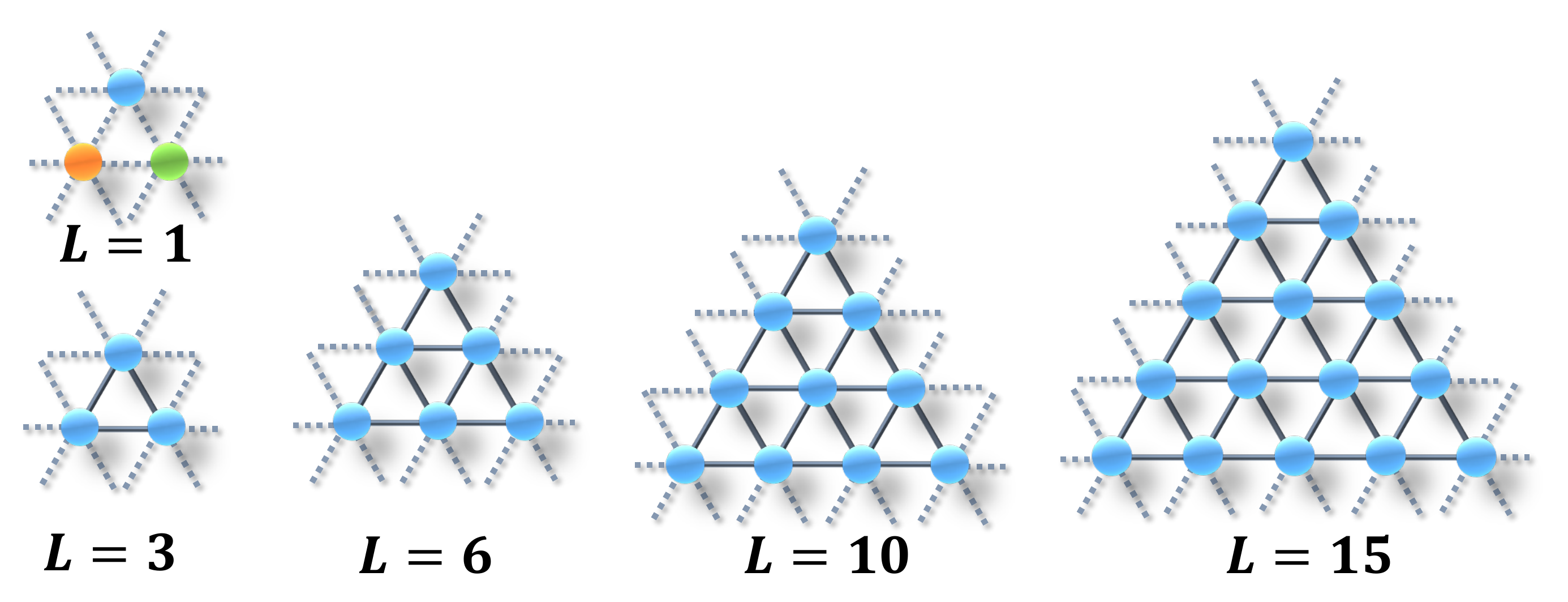}
\caption{\label{clusters}(a) The units in the Gutzwiller factorization of the full lattice. We label the clusters by the number of involved sites $L$ in it. In the standard ($L=1$) mean-field approximation, the interactions between any site to its nearest-neighbors are treated as an effective field (the dashed bonds), generated by the neighboring sites, imposing on it. In particular, we divide the full lattice into three sublattices labeled by $A$, $B$, and $C$ (in different colors). For clusters of sizes $L\ge3$, the interactions between the nearest-neighboring sites inside the cluster (the solid bonds) are treated exactly. The interactions between the sites on the boundaries to its neighboring sites outside the cluster are treated by the effective fields.}
\end{figure}

We start with the mean-field approximation and will refine the mean-field results by taking the short-range correlations into account in next section. Generally, the mean-field approximation is based on the Gutzwiller factorization by representing the global density matrix as a tensor product of the state of each site, $\rho = \bigotimes_j\rho_j$. In order to reveal the antiferromagnetic nature we divide the whole lattice into three sublattices $A$, $B$ and $C$ as shown in Fig. \ref{clusters}. Assuming that all the sites belonging to the same sublattice are identical, we may obtain the single-site MF master equation for each sublattice as the followings,
\begin{equation}
\frac{d\rho_{j}}{dt}=-i[H^{\text{MF}}_j,\rho_{j}] + \frac{\gamma}{2}\left[2\sigma_j^{-}\rho_{j}\sigma_j^{+}-\{\sigma_j^{+}\sigma_j^{-},\rho_{j}\}\right],
\label{eq:mfmasterequation}
\end{equation}
where $j = A$, $B$ and $C$ denotes the sublattice. The MF Hamiltonian for sublattice $j$ is given by
\begin{equation}
H^{\text{MF}}_j = \sum_{\alpha = x,y,z}{\sum_{k\neq j}{J_\alpha\langle\sigma^z_k\rangle\sigma^\alpha_j}},
\end{equation}
where $\langle\sigma^{\alpha}_{k}\rangle$ represent the expectation value of the spin operator for the sublattice $k$. The MF master equations can be reexpressed by the system of Bloch equations as
\begin{equation}
\begin{aligned}
\frac{d\langle \sigma^x_{j}\rangle}{dt}&=2\sum_{k\neq j}[J_y\langle\sigma^z_{j}\rangle\langle\sigma^y_{k}\rangle-J_z\langle\sigma^y_{j}\rangle\langle\sigma^z_{k}\rangle]-\frac{\gamma}{2}\langle\sigma^x_j\rangle,\\
\frac{d\langle \sigma^y_{j}\rangle}{dt}&=2\sum_{k\neq j}[J_z\langle\sigma^x_{j}\rangle\langle\sigma^z_{k}\rangle-J_x\langle\sigma^z_{j}\rangle\langle\sigma^x_{k}\rangle]-\frac{\gamma}{2}\langle\sigma^y_j\rangle,\\
\frac{d\langle \sigma^z_{j}\rangle}{dt}&=2\sum_{k\neq j}[J_x\langle\sigma^y_{j}\rangle\langle\sigma^x_{k}\rangle - J_y\langle\sigma^x_{j}\rangle\langle\sigma^y_{k}\rangle]-\gamma(\langle\sigma^z_j\rangle + 1).
\end{aligned}
\label{blocheqn}
\end{equation}

We note that actually there are nine nonlinear differential equations in Eq. (\ref{blocheqn}) since there are three sublattices. For each sublattice, its nearest neighbours consist of the other two sublattices. The steady-state magnetizations of the system are given by the stable fixed points of the nonlinear equations. We numerically find the fixed points by setting Eqs. (6) to be zero and check their stabilities through the eigenvalues of the Jacobian for each fixed point. If the real parts of all the eigenvalues are negative the fixed point is stable, otherwise it is unstable. The presence of two conjugated pure imaginary eigenvalues implies the appearance of limit cycle. In the rest of this paper, we work in units of $\gamma$.

On the triangular lattice, due to the presence of geometric frustration, the antiferromagnetic phase is much richer than the counterpart in the square lattice. We find that the steady-state is possible to enter into two types of antiferromagnetic phase:

(a) {\it Biantiferromagnetic (BAF) phase}. The BAF phase corresponds to the case that two sublattices have the same nonzero magnetization in the $x$-$y$ plane but different from the third one, i.e. $\langle\sigma^{\alpha}_{A}\rangle_{\text{ss}} = \langle\sigma^{\alpha}_{B}\rangle_{\text{ss}} \neq \langle\sigma^{\alpha}_{C}\rangle_{\text{ss}}$ ($\alpha=x,y$).

(b) {\it Triantiferromagnetic (TAF) phase}. The TAF phase corresponds to the case that the steady-state magnetization of the three sublattices are completely unequal, i.e.  $\langle\sigma^{\alpha}_{A}\rangle_{\text{ss}} \neq \langle\sigma^{\alpha}_{B}\rangle_{\text{ss}} \neq \langle\sigma^{\alpha}_{C}\rangle_{\text{ss}}$ ($\alpha=x,y$).

\begin{figure}[!htbp]
\includegraphics[width=1\linewidth]{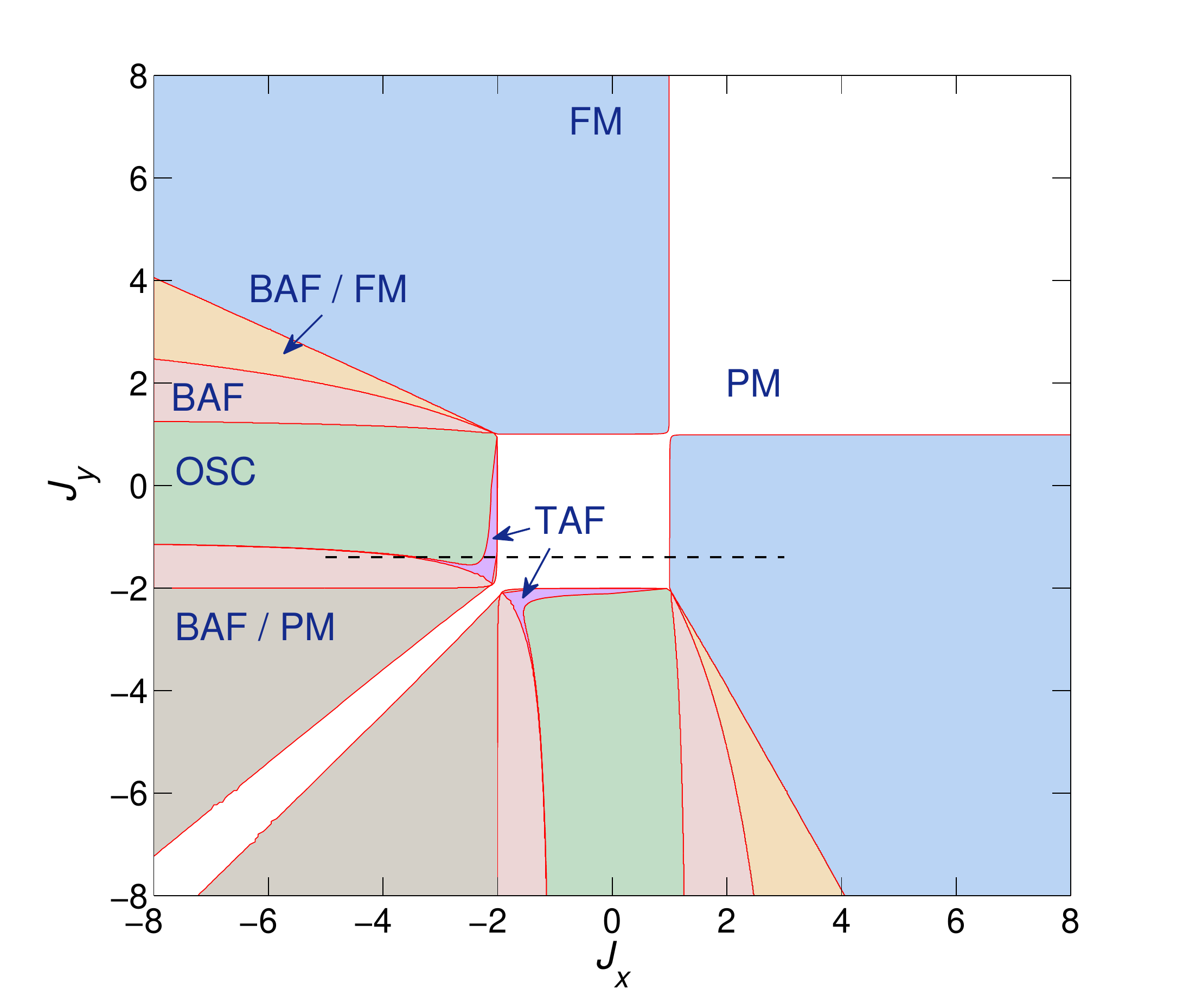}
\caption{\label{mf_pd} The mean-field phase diagram in the $J_{x}$-$J_{y}$ plane. The PM phase (white region) is uniform and preserves the $\mathbb{Z}_2$ symmetry. The FM phase is also uniform but the $\mathbb{Z}_2$ symmetry is broken. The BAF and TAF phases are non-uniform.
The OSC phase breaks the time-translation invariance resulting a limit cycle in the long-time limit. There two bistable regions regarding the BAF/FM and BAF/PM in which the asymptotic steady states depend on the initial state. The parameter is chosen as $J_z = 1$. The horizontal dashed line marks the cut along $J_y = -1.4$ which will be scrutinized in Fig. \ref{mf_magn}.}
\end{figure}

We show the mean-field phase diagram in Fig. \ref{mf_pd} which is symmetrized about $J_x = J_y$. The order parameter is chosen as the steady-state magnetization along the $x$-direction $\langle \sigma^x_j\rangle^{\text{mf}}_{\text{ss}}$. Basically, there are five steady-state phases in the $J_x$-$J_y$ plane corresponding to the PM, FM, BAF, TAF, and oscillatory (OSC) phases and two bistable regions with respect to the BAF/PM and BAF/FM phases. In the bistable region, the system evolves to two different phases depending on the initial state.

It is possible to analytically determine the critical points of the PM-FM or PM-TAF phase transitions. For the PM-FM phase transition, the critical points are given
\begin{equation}
J_{x,y}^c = J_z + \frac{1}{16\mathfrak{z}^2(J_z-J_{y,x})},
\end{equation}
while for the PM-TAF phase transtion, the critical points are given by
\begin{equation}
J_{x,y}^c = -2J_z - \frac{1}{4\mathfrak{z}^2(J_{y,x}+2J_z)},
\end{equation}
where $\mathfrak{z}=6$ is the coordinate number of the triangular lattice.

\begin{figure}
\centering
\begin{minipage}[b]{1\linewidth}
\includegraphics[width=1\linewidth]{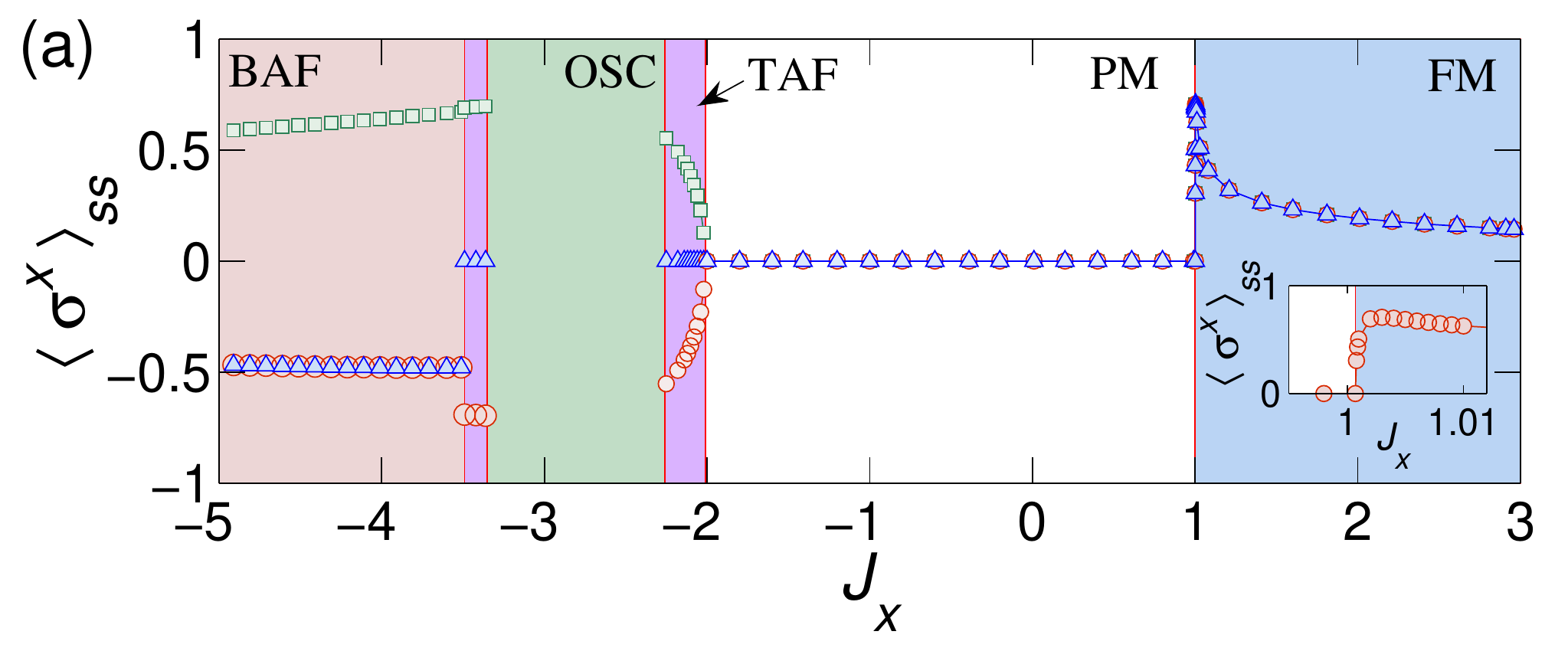} \\
\includegraphics[width=1\linewidth]{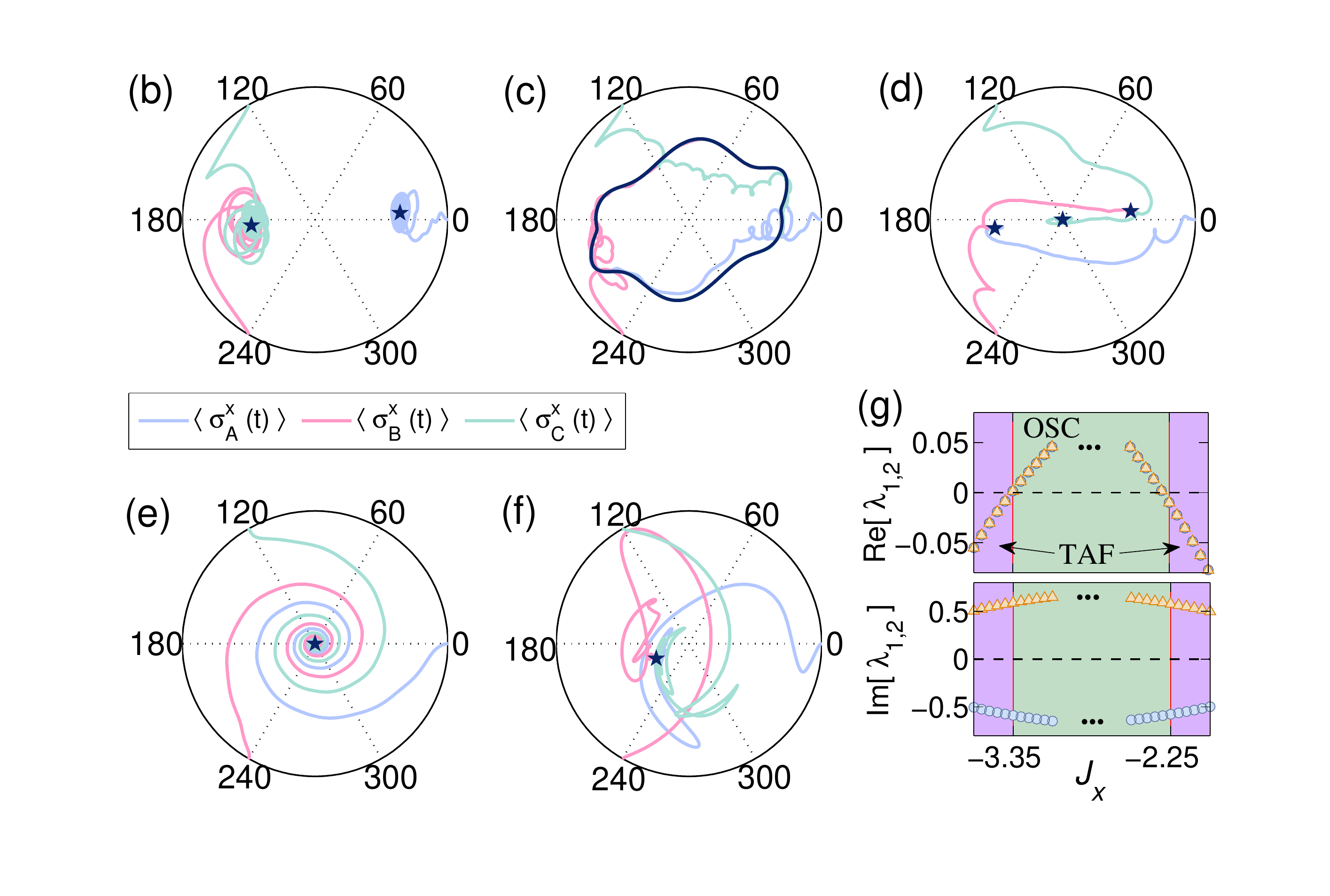}
\caption{\label{mf_magn}(a) The mean-field steady-state magnetization along $x$-direction of three sublattices (indicated by various symbols) for a horizontal cut of Fig. \ref{mf_pd} along $J_y = -1.4$. Particularly, in the OSC region the data are lacking since the system never reaches to the asymptotic state in the long-time limit. In the inset we show a zoom-in near the transition between the PM and FM phases. (b)-(f) The time-evolution of trajectories of the magnetizations of each sublattice in the equatorial plane of the Bloch sphere for $J_x = -4$, $-2.8$, $-2.2$, $-1.5$ and $1.5$ which correspond to the BAF, OSC, TAF, PM and FM regions respectively. The initial state is chosen as the $120^{\circ}$ state. The attractors are the stable fixed points (indicated by the stars) in panels (b), (d)-(f) and limit cycle (indicated by the black solid line) in panel (c). (g) The real and imaginary parts of $\lambda_1$ and $\lambda_2$, which are the eigenvalues of the Jacobian associated with the TAF solutions to Eq. (\ref{blocheqn}), with most (triangles) and second (circles) largest real parts as functions of $J_x$. }
\end{minipage}
\end{figure}

We deliberately consider a horizontal cut of Fig. \ref{mf_pd} at $J_y = -1.4$ since this line passes through the typical phases of the present model. The steady-state magnetization is shown in Fig. \ref{mf_magn}(a). It can be seen that the PM-FM and TAF-FM transitions are continuous and, close to the transition, the growth of the order parameter is described by $\langle\sigma^x\rangle_{\text{ss}}^{\text{mf}}\sim (J_{x} - J_{x}^c)^{1/2}$ where $J_x^c$ is the transition point. In the PM phase, all spins are uniformly pointing down to the $z$-direction and the steady state is $|\downarrow\downarrow\downarrow...\rangle$. Starting from the PM phase, on the one hand, as the coupling strength $J_x$ increasing the system enters to the FM phase. In the FM phase, all the sublattices acquire the same nonzero magnetization along the $x$-$y$ plane. The system is still uniform. On the other hand, as $J_x$ decreasing, the steady-state magnetizations of two sublattices deviate from the $|\downarrow\rangle$ state and the magnetization component on the $x$-$y$ plane grow along opposite directions while the third one stays in the $|\downarrow\rangle$ state, indicating that the system enters into the TAF phase.

In an intermediate region $-3.35 \le J_x \le -2.25$, none of the asymptotic steady-state phases are stable and the limit cycle appears. This is supported by the behavior the eigenvalues of the Jacobian for the TAF fixed point. From Fig. \ref{mf_magn}(g), one can see that, starting from the TAF phase, a pair of complex conjugate eigenvalues cross the imaginary axis (namely two pure imaginary eigenvalues present) at the the transition points $J_x=-3.35$ and $-2.25$ signaling the appearance of the limit cycle.

As $J_x$ continues to decrease, the TAF phase is replaced by the BAF phase at $J_x \le -3.49$ and the magnetization of all the three sublattices have nonzero component in the equatorial plane. The continuity of the PM-FM and PM-TAF transitions can be seen from the value of $\langle\sigma^x\rangle^{\text{mf}}_{\text{ss}}$ in the vicinity of the critical points. Additionally, the transition between the nonuniform BAF and TAF phases is discontinuous characterized by sudden change of the order parameter.

In order to give an intuitive description, we show the time-evolution of the magnetization on the equatorial plane of Bloch sphere for each phase in Fig. \ref{mf_magn}(b)-(f). We initialize the sublattice by the $120^\circ$ state on the equatorial plane of the Bloch sphere, i.e. $|\psi_A(0)\rangle = (|\uparrow\rangle+|\downarrow\rangle)/\sqrt{2}$, $|\psi_B(0)\rangle = (|\uparrow\rangle+e^{i2\pi/3}|\downarrow\rangle)/\sqrt{2}$ and $|\psi_C(0)\rangle = (|\uparrow\rangle+e^{-i2\pi/3}|\downarrow\rangle)/\sqrt{2}$. It is found that the system always falls onto the stable fixed points. For the uniform phases, there is only one fixed point. More specifically, for the PM phase the fixed point locates on the origin while for the FM phase the fixed point deviates from the origin, implying the preservation and breakdown of the $\mathbb{Z}_2$-symmetry. For the nonuniform phases, there are two and three fixed points for the BAF and TAF phases, respectively. Interestingly, there exists an OSC phase in which the long-time limit trajectory is a limit cycle breaking the time-translation symmetry. The presence of the OSC phase in the square lattice has been ruled out in the literature \cite{owen2018}. Here, the presence of oscillatory phase in frustrated lattice is probably an artifact of the sublattice factorization of the full lattice. As will be seen in next section, the OSC region is replaced by the time-translation invariant phases in the thermodynamic limit.

\section{Beyond single-site mean-field approximation}
\label{section:beyond mean-field approximation}
In this section we will investigate the steady-state phases beyond mean-field approximation. It has been proven that the short-range correlations play crucial role in determining the structure of the steady-state phase diagram. Here, we systematically include the short-range correlations in the analysis by exploiting the CMF approach. The idea of CMF approach is to decouple the Hamiltonian of the full lattice into Hamiltonians of identical clusters consists of contiguous sites. The interactions inside the cluster are faithfully described while the interactions between sites belonging neighbouring clusters are effectively treated by mean-field terms. More details about the CMF approach can be found in Ref. \cite{jin2016}.

For the present model, we perform the CMF studies with a series of triangular clusters up to size $L=15$ as shown in Fig. \ref{clusters}. For $L = 3, 6$ and $10$ we simulate the evolution of the CMF master equation with a fourth-order Runge-Kutta algorithm while for $L=15$ we resort to the quantum trajectory. In this work, the data obtained by quantum trajectory are averaging over $500$ trajectories.

\subsection{The PM-FM transition}
\begin{figure}[!htbp]
\includegraphics[width=1\linewidth]{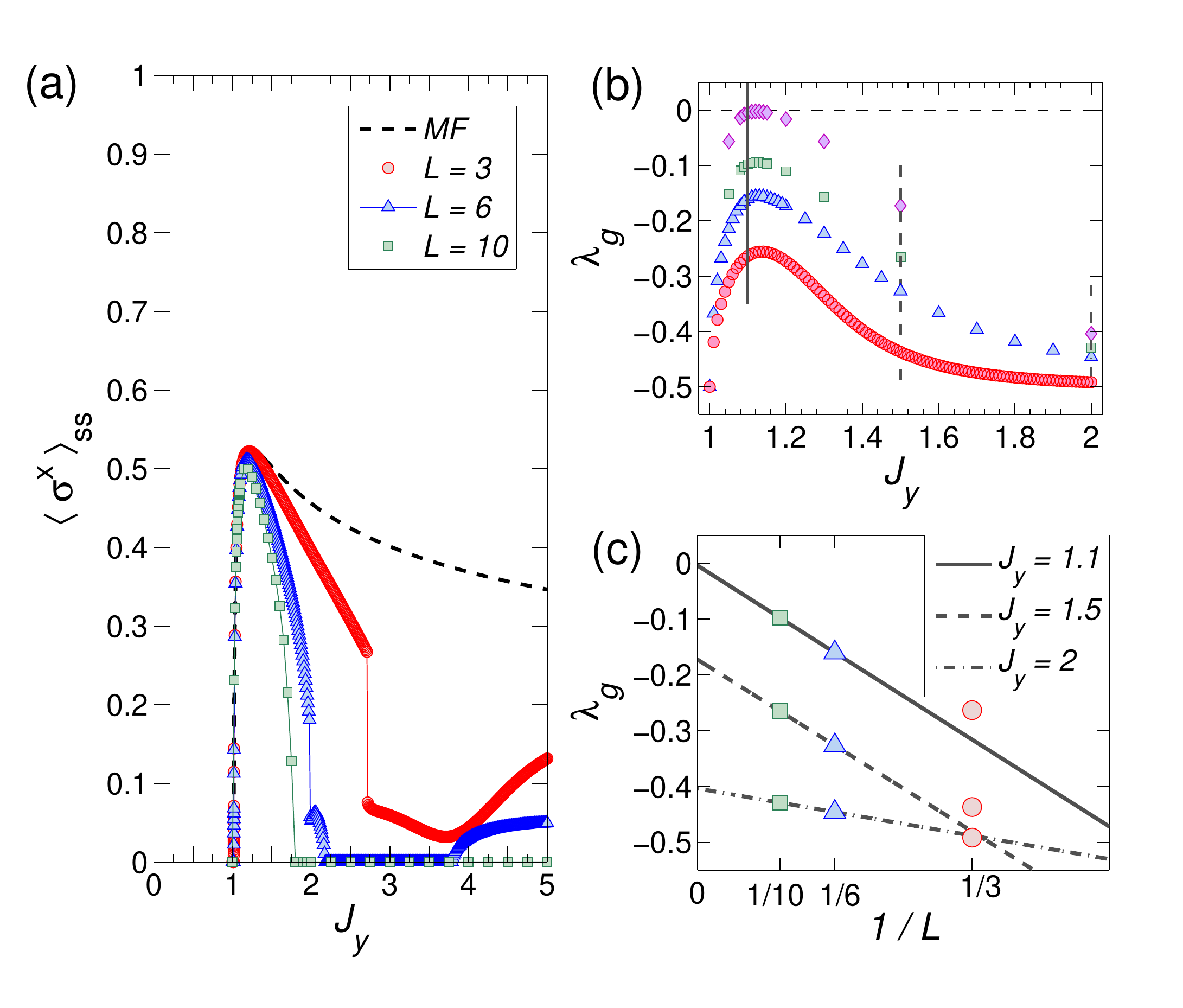}
\caption{\label{fig4} (a) The steady-state magnetization $\langle\sigma^x\rangle_{\text{ss}}$ as a function of $J_y$. The dashed black line indicates the MF solutions. Symbols are the CMF results for clusters of various size as indicated in the legend (applied to all the panels). (b) The real part of the Liouvillian gap $\lambda_g$ as a function of $J_y$ for various clusters with open boundary condition (without mean field). The diamond indicates the extrapolated gap in the thermodynamic limit ($L\rightarrow\infty$). The vertical dark lines are guide for eyes. (c) The linear fit of $\lambda_g$ to $1/L$. Other parameters are chosen as $J_x = 0.9$ and $J_z = 1$.}
\end{figure}

We first analyze the steady-state PM-FM phase transition along a vertical cut along $J_x=0.9$ in Fig. \ref{mf_pd}. We employ the expectation value of the magnetization of each site $\langle\sigma^x\rangle$ as the order parameter. The resulting magnetization is shown in Fig. \ref{fig4} with different symbols denoting various clusters. Recall that the MF approximation predicts that the FM phase exists for $J_y \ge J^c_y\approx1.0174$. Here we find that the FM region is fragile to the inclusion of short-range correlations in the analysis. As the size of cluster systematically increasing, the FM phase shrinks down to a closed region. At the level of $L=3$, the steady-state magnetization changes abruptly at $J_y\approx 2.55$ and is strongly suppressed for $J_y>2.55$; at the level of $L=6$, in addition to the abrupt change, the magnetization vanishes in an intermediate region of $2.18\lesssim J_y\lesssim 3.79$ and revives for large $J_y\ge3.79$. Eventually, at $L=10$ level, the FM phase only survives in a closed region of $1.0175\lesssim J_y\lesssim1.75$. We conclude that the PM-FM phase transition for the chosen parameter is replaced by the successive PM-FM-PM transitions. Such a behavior is similar to the counterpart in the square lattice.

The fact that the ordered phase is confined to a closed region is also revealed by the Liouvillian spectrum \cite{minganti2018}.
It has been proven that the real part of the eigenvalues of Liouvillian superoperator $\mathcal{L}$ is always non-positive \cite{breuer_book}. The Liouvillian gap $\lambda_g$ is defined as the largest negative real part of the eigenvalues, which is also called the asymptotic decay rate since it describes the slowest relaxation scale toward the steady state \cite{kessler2012}. In Fig. \ref{fig4}(b), it is shown the Liouvillian gap $\lambda_g$  for various lattices. A linear fitting of these data with $1/L$ extrapolates the Liouvillian gap in the thermodynamic limit, as indicated by the diamonds in Fig. \ref{fig4}(b). One can see that the Liouvillian gap is closed over an intermediate region of $J_y$ implying the existence of the ordered FM phase. We show the scaling of Liouvillian gaps for three different values of $J_y$ in Fig. \ref{fig4}(c). For $J_y=2$, the Liouvillian gap is finite in the thermodynamic limit corresponding to the reentrance to the PM region. For $J_y=1.1$, which locates deeply in the FM region, the close of Liouvillian gap in the thermodynamic limit has been observed. While for $J_y=1.5$, which is close to the transition, extra data of larger size ($L>10$) cluster should be considered to get more faithful result in the thermodynamics limit since the correlation length diverges.

\subsection{The antiferromagnetic phases}
\begin{figure}[h]
\includegraphics[width=0.9\linewidth]{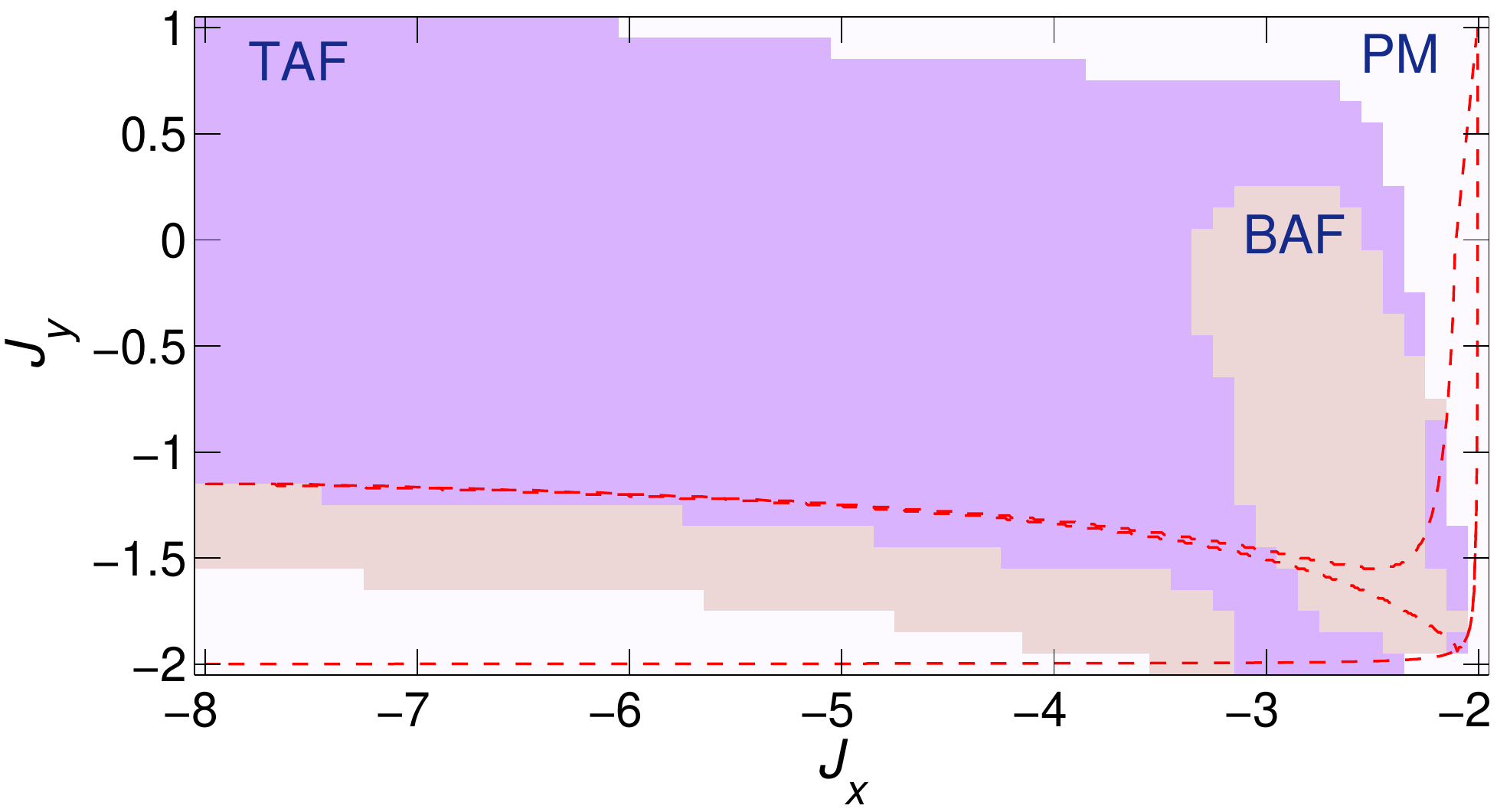}
\caption{\label{pdT06} The CMF phase diagram in the $J_x$-$J_y$ plane for $L=6$. Compared with the MF phase diagram, the TAF phase extends while the OSC phase disappears. The boundaries between the PM and the non-uniform (BAF and TAF) phases are deformed. A new BAF region emerges inside the TAF phase. The parameter is chosen as $J_z=1$. For comparison, the MF phase boundaries extracted from Fig. \ref{mf_pd} are presented by the dashed lines.}
\end{figure}

We now investigate the antiferromagnetic phases by focusing on the region of $-8\le J_x\le -2$ and $-2\le J_y\le1$. Under the single-site mean-field approximation, such a region contains the OSC, TAF and BAF phases. In Fig. \ref{pdT06} we show the CMF phase diagram with the cluster of $L=6$. Compared with Fig. \ref{mf_pd}, firstly, we see that the TAF phase expands and replaces the OSC phase. Secondly, the inclusion of the effects of correlations modifies the boundary between the PM and nonuniform (BAF and TAF) phases; the area of nonuniform phases is reduced. Moreover, the BAF/PM bistable region does not exist any more. Thirdly, it is interesting that an island of BAF emerges in an intermediated region of $-3.2\lesssim J_x \lesssim -2$ and $-2\lesssim J_y\lesssim 0.25$. The emerged BAF region is completed missed by the single-site MF approximation and survives in the thermodynamic limit.

In order to show various phases, we employ the order parameter ${\cal O}_{\text{AF}}=\frac{1}{L}\sum_{j=1}^L{|\langle\sigma^x_j\rangle_{\text{ss}}|}$ to distinguish the PM and the (nonuniform) antiferromagnetic phases. The vanishing value of ${\cal O}_{\text{AF}}$ indicates that the system is in the PM phase. Moreover, we define another order parameter ${\cal O}_{\text{nTAF}}=\frac{1}{L}|\sum_{j=1}^{L}{\langle\sigma_j^x\rangle_{\text{ss}}}|$ to distinguish the TAF phase and other antiferromagnetic phases. Such order parameter measures the net magnetization on the $x$-$y$ plane and thus vanishes for the TAF phase. We focus on a horizontal line of Fig. \ref{pdT06} along $J_y = -1.7$ and show the order parameters as functions of $J_x$ for CMF computation with clusters of different sizes.

\begin{figure}[!htbp]
\includegraphics[width=1\linewidth]{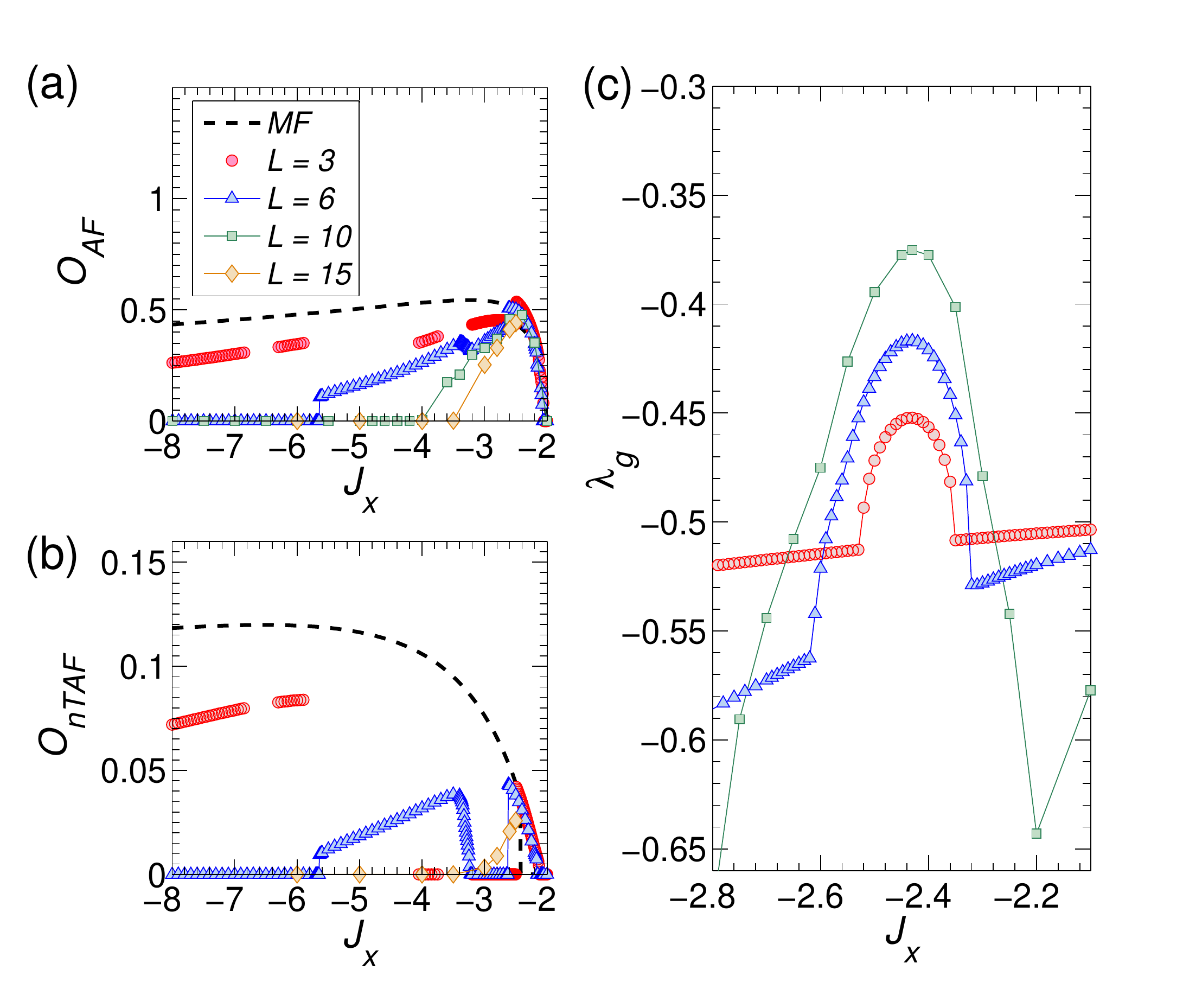}
\caption{The order parameters ${\cal O}_{\text{AF}}$ (a), ${\cal O}_{\text{nTAF}}$ (b) and Liouvillian gap (c) as functions of $J_x$ for various size of clusters (denoted by different symbols). The parameters are chosen as $J_y = -1.7$ and $J_z = 1$.}
\label{AForder}
\end{figure}

In Fig. \ref{AForder} we show the order parameters as a function $J_x$ for CMF approximation with clusters of different sizes. From Fig. \ref{AForder}(a), one can see that as the size of cluster increases the value of ${\cal O}_{\text{AF}}$ is suppressed for smaller $J_x$ and eventually vanishes for $J_x\lesssim-3.5$ at $L=15$. This implies that the antiferromagnetic phases shrink into a closed region in the presence of correlations which is similar to that happens in FM phase. Now we analyze the interior of the antiferromagnetic phase by virtue of the order parameter ${\cal O}_{\text{nTAF}}$. In Fig. \ref{AForder}(b), we show ${\cal O}_{\text{nTAF}}$ as a function of $J_x$. Indeed, one can see that the BAF appears in $-3\lesssim J_x\lesssim -2.2$ with nonzero values of ${\cal O}_{\text{nTAF}}$, while the TAF exists in $-3.5< J_x < -3$ with vanishing ${\cal O}_{\text{nTAF}}$ but nonzero ${\cal O}_{\text{AF}}$. The emergence of the BAF phase is also witnessed by the Liouvillian gap as shown in Fig. \ref{AForder}. A peak of the Liouvillian gap abruptly appears in a finite domain of the parameter $J_x$. As the size of the finite lattice increasing, the peak raises and broadens implying the appearance of symmetry-breaking BAF phases. In addition, the Liouvillian spectrum for small $J_x$ remains gapped (not shown).

\subsection{Spin-density-wave phase}
Apart from the FM, BAF and TAF phases, the system may evolve to a spatially modulated state with incommensurate order. This is uncovered by the linear stability analysis on the product spin-down state $\rho_{\downarrow}=\bigotimes_{j} \rho_{j,\downarrow}$ where $\rho_{j,\downarrow}=|\downarrow_j\rangle\langle\downarrow_j|$. It is easy to see that $\rho_{j,\downarrow}$ is always a steady-state solution to the MF master equation Eq. (\ref{eq:mfmasterequation}). The idea of linear stability analysis is to introduce local perturbations $\delta\rho_j$ to modify the MF steady state by $\rho_{\downarrow}\rightarrow \bigotimes_j\left(\rho_{j,\downarrow}+\delta\rho_j\right)$, and check how the perturbations evolve with time. If the perturbations are suppressed in the long-time limit, the factorized density matrix is stable; otherwise, new ordering may be established in the system.

\begin{figure}[htbp]
\includegraphics[width=1\linewidth]{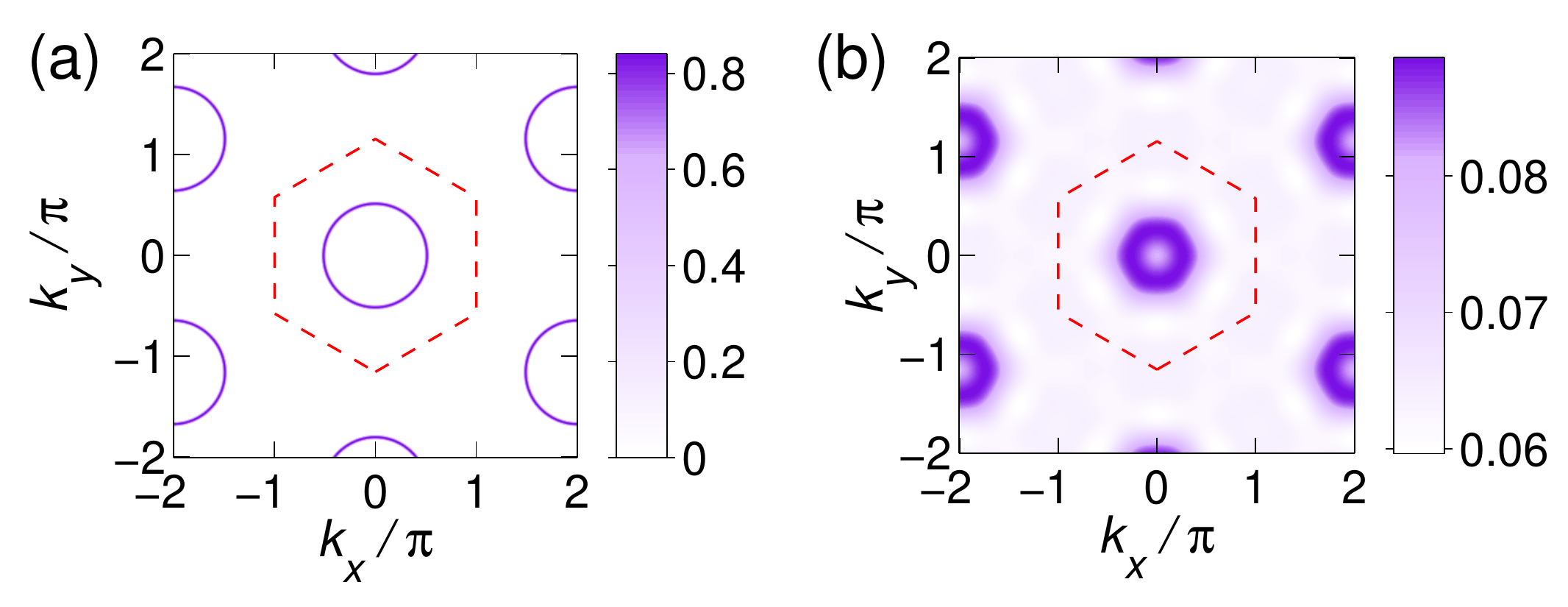}
\caption{ (a) Real part of the most unstable eigenvalue as a function $k_x$ and $k_y$.  (b) The steady-state spin-structure factor in the momentum space. The computation is implemented on $L=15$ triangular cluster with open boundary condition without mean fields. The parameters are chosen as $J_x = 2$, $J_y=2.5$ and $J_z = 1$.}
\label{ssf_T15}
\end{figure}

It is convenient to expand the perturbations in terms of plane waves
\begin{equation}
\delta\rho_j=\sum_{\textbf{k}}{e^{-i\textbf{k}\cdot\textbf{r}_j}\delta\rho_{\textbf{k}}}.
\end{equation}

Thus, the equation of motion for the perturbation $\delta\rho_{\textbf{k}}$ reads
\begin{equation}
\partial_t{\delta\rho_{\textbf{k}}}=\delta{\cal L}\left[\rho_{\text{ss}}\right]+{\cal L}\left[\delta\rho_{\textbf{k}}\right]+{\cal D}\left[\delta\rho_{\textbf{k}}\right],
\label{LSA}
\end{equation}
where $\delta{\cal L}\left[\rho_{\text{ss}}\right]=-i\sum_{\textbf{e}_{n},\alpha}{ J_\alpha[e^{i\textbf{k}\cdot\textbf{e}_n}\text{tr}(\sigma^\alpha\delta \rho_{\textbf{k}})\sigma^\alpha,\rho_{\text{ss}}]}$, and ${\cal L}\left[\delta\rho_{\textbf{k}}\right]=-i\sum_{\alpha}{[J_\alpha\text{tr}(\sigma^\alpha\rho_\text{ss})\sigma^\alpha,\delta\rho_{\textbf{k}}]}$ ($\alpha=x,y,z$). The vectors
%$\vec{e}_n\in[\pm\vec{e}_x,\pm1/2\vec{e}_x\pm\sqrt{3}/2\vec{e}_y]$
$\textbf{e}_n=\cos{(n\pi/3)}\textbf{e}_x+\sin{(n\pi/3)}\textbf{e}_y$ ($n=0,1,...,5$)
denote the directions to the neighboring sites. Eq. (\ref{LSA}) can be recast as $\partial_t{\delta\rho_{\textbf{k}}}={\cal M}[\delta\rho_{\textbf{k}}]$ where ${\cal M}$ is a superoperator acting on the perturbation. The stability of the relevant steady state is determined by the spectrum of ${\cal M}$. If the real parts of all the eigenvalues of ${\cal M}$ are negative, the steady state is stable; otherwise it is unstable to perturbations of wave vector $\textbf{k}=(k_x,k_y)$. Substituting the steady state $\rho_{\downarrow}$ into Eq. (\ref{LSA}), we find that the PM phase is unstable when
\begin{equation}
\left(J_{x}\sum_{m=1}^{\mathfrak{z}}e^{i\cdot \textbf{k}_{m}}-\mathfrak{z}J_{z}\right)\left(J_{y}\sum_{m=1}^{\mathfrak{z}}e^{i\cdot \textbf{k}_{m}}-\mathfrak{z}J_{z}\right)<-\frac{\gamma^2}{16}.
\label{Eq.unstablecondition}
\end{equation}

In Fig. \ref{ssf_T15}(a) we show the largest real parts of the eigenvalues, associated to the most unstable wave vector, of ${\cal M}$ in momentum space for $J_x=2$ and $J_y=2.5$. We can see a finite-momentum instability in the first Brillouin zone. In this case the PM phase is mostly unstable to momenta $|\textbf{k}|\approx 0.51$ implying an spin-density wave of wavelength $\lambda\approx 3.92$ may be established. To corroborate the appearance of the SDW phase, we calculate the steady-state spin-structure factor $S_{\text{ss}}^{xx}(\textbf{k})$ which is defined as
\begin{equation}
S_{\text{ss}}^{xx}\left(\textbf{k}\right)=\frac{1}{L^2}\sum_{j,l=1}^{L}{e^{-i\textbf{k}\cdot(\textbf{r}_j-\textbf{r}_l)}\langle\sigma^x_j\sigma^x_l\rangle_{\text{ss}}}.
\end{equation}
A nonzero value of $S_{\text{ss}}^{xx}\left(\textbf{k}\right)$ indicates the stabilization of spin-density wave with momentum $\textbf{k}$. In Fig. \ref{ssf_T15}(b) we show $S_{\text{ss}}^{xx}\left(\textbf{k}\right)$ in the momentum space for a cluster of $L=15$ (without mean fields) with open boundary condition. The SDW feature is explicitly signaled by the ring-shape of $S_{\text{ss}}^{xx}\left(\textbf{k}\right)$. The mismatch of the wavelengths in Fig. \ref{ssf_T15}(a) and (b) is due to the finite-size effect.

\section{Summary}
\label{section:conclusions}
In summary, we have investigated the steady-state phase diagram of dissipative spin-1/2 XYZ model on a triangular lattice. We employed the mean-field type approximations based on the Gutzwiller factorization of the global density matrix to decouple the full master equation. In the single-site MF analysis, we numerically solved self-consistent Bloch equations and check the stabilities for the fixed points. Due to the geometric frustrations in the lattice, we found that the MF steady-state phase diagram is rather rich including the PM, FM, BAF, TAF and OSC phases.

In order to confirm the existence of the various steady-state phases in the thermodynamic limit, we gradually included the short-range correlations in the analysis by enlarging the sizes of clusters. We demonstrated that FM phase shrinks into a closed region which is supported by the closing of Liouvillian gap in the $\mathbb{Z}_2$ symmetry-broken phase. We also showed that the antiferromagnetic region in the mean-field approximation changed dramatically: first the OSC phase is erased and replaced by the TAF phases; second a new region of BAF emerges in an intermediate region inside the TAF phase. Disappearance of the limit cycles, when the short-range correlations are considered in the analysis, indicates that the OSC phase is an artifact of the MF treatment. Moreover, we found the evidence of the SDW phase, which is missed by the MF approximation, by performing the linear stability analysis. The feature of SDW phase was revealed by the spin-structure factor of a finite-size cluster. To further characterize the steady-state phase transitions, the quantum fisher information is proved to be a potential indicator \cite{braun2018,marzolino2017}.

Finally, we note that identifying the steady-state phases of open quantum many-body system is rather a demanding task. Due to the exponential growth of the many-body Hilbert space, we stopped our CMF computation at the cluster with $L=15$. To exactly locate the boundary of phases and extract the critical exponents, further analysis on larger size clusters are required. Along this line, the combination of the CMF approximation with other algorithms, such as corner-space renormalization \cite{finazzi2015}, matrix-product-operator approach \cite{mascarenhas2015}, and neural networks \cite{yoshioka2019,nagy2019,hartmann2019,vicentini2019} should be helpful.

\section*{ACKNOWLEDGMENTS}
\label{setion:acknowlegments}
This work is supported by National Natural Science Foundation of China under
Grant No. 11975064 and No. 11775040, and the Fundamental Research Funds for the Central Universities No. DUT19LK13.

\end{document}